\documentclass[aps,prl,twocolumn,superscriptaddress,amssymb]{revtex4-1}

\usepackage{graphicx,amsmath}

\begin{document}

\title{
Class $\mathcal S$ Anomalies from M-theory Inflow
}

\author{Ibrahima Bah}
\author{Federico Bonetti}
\affiliation{Department of Physics and Astronomy, Johns Hopkins University, 3400 North Charles Street, Baltimore, MD 21218, USA}
\author{Ruben Minasian}
\affiliation{Institut de Physique Th\'{e}orique, Universit\'{e} Paris Saclay, CNRS, CEA,  F-91191, Gif-sur-Yvette, France}
\author{Emily Nardoni} 
\affiliation{Mani L. Bhaumik Institute for Theoretical Physics, Department of Physics and Astronomy, University of California, Los Angeles,  CA 90095, USA}

\begin{abstract}

We present a first principles
derivation of the anomaly polynomials of 4d $\mathcal N = 2$
class $\mathcal S$ theories of type $A_{N-1}$ with arbitrary regular punctures,
using anomaly inflow in the corresponding M-theory setup
with $N$ M5-branes wrapping a punctured Riemann surface.
The labeling of punctures in our approach follows entirely
from the analysis of the 11d geometry and $G_4$ flux.
We highlight the applications of the inflow method 
to the AdS/CFT correspondence.

\end{abstract}


\maketitle

\section{Introduction  \label{intro}}

't Hooft anomalies are measures of degrees of freedom of quantum systems that are preserved under renormalization group flow.  Thus, anomalies provide powerful tools for exploring phases and non-perturbative regimes of quantum theories. 

In the last ten years, a new approach to studying quantum field theories (QFTs) has emerged with the discovery of $\mathcal{N}=2$ class $\mathcal{S}$ superconformal field theories (SCFTs) \cite{Gaiotto:2009we,Gaiotto:2009hg}, where
a large class of
 4d $\mathcal{N}=2$ SCFTs are geometrically defined from reductions of 6d $(2,0)$ SCFTs on punctured Riemann surfaces.  A choice of 6d SCFT and boundary data at the punctures completely specify a 4d SCFT and its various protected sectors.  A typical theory in this class is non-Lagrangian and strongly coupled, and yet it can be analyzed from the geometric construction.  The approach of the class $\mathcal{S}$ program has been generalized and adopted for studying SCFTs in different dimensions with varying amount of supersymmetry.  The geometrization program has become a standard tool in the study of QFTs.

A key feature of the class $\mathcal{S}$ program is the richness of the variety of punctures on the Riemann surface. The anomalies of
$\mathcal N = 2$
 class $\mathcal{S}$ SCFTs in the presence of regular punctures have been indirectly obtained from field theoretic arguments \cite{Gaiotto:2009gz,Chacaltana:2012zy,Tachikawa:2015bga}. However, a direct derivation of the anomalies from the geometric definition of class $\mathcal{S}$ SCFTs is lacking. In this letter we use anomaly inflow in M-theory to provide a first principles derivation,
building on \cite{Bah:2018gwc}. 
Our procedure can be generalized to obtain the anomalies of other classes of SCFTs with geometric descriptions.
Further, 
our prescription  suggests a method for extracting the
exact anomalies of a holographic SCFT from its 
gravity dual.

The 't Hooft anomalies of a $d$-dimensional QFT are neatly encoded in the $(d+2)$-form anomaly polynomial. 
In this letter we derive the  
anomaly polynomials of 4d $\mathcal{N}=2$ class $\mathcal{S}$ SCFTs with regular punctures engineered from the 6d $(2,0)$ $A_{N-1}$  SCFTs.  First, we describe the relevant geometric setup from a stack of $N$ M5-branes in M-theory, and the inflow procedure.  Then we provide a novel description of 
the
boundary data at punctures in terms of the four-form flux of M-theory.  Finally, we compute the anomaly polynomial and discuss its implications for holography.  
A companion paper \cite{Bah:2018bn} to this letter
contains more complete derivations and a broader study
of the results and their implications.

\section{Setup and Inflow}

A 4d $\mathcal N = 2$  class $\mathcal S$   theory
of type $A_{N-1}$ is engineered in M-theory
by taking the low-energy limit of a configuration with 
$N$ coincident M5-branes wrapping a punctured Riemann surface.
Let $W_6$ denote the 6d worldvolume
of the M5-brane stack inside the ambient 11d space $M_{11}$.
The normal bundle to $W_6$,  denoted $NW_6$,~encodes the five transverse directions to
the stack and 
generically has structure group $SO(5)$.
We study the case 
 $W_6 = M_4 \times \Sigma_{g,n}$,
where $M_4$ is external spacetime
and $\Sigma_{g,n}$ is a Riemann surface of genus $g$ with $n$ punctures.

We are interested in setups that 
preserve 
4d  $\mathcal N = 2$ supersymmetry
(for $M_4 = \mathbb R^{1,3}$).
In this case, 
the structure group
of   $NW_6$ reduces
from $SO(5)$ to $SO(2) \times SO(3)$,
and correspondingly  $NW_6$
decomposes as $NW_6 = N_{SO(3)} \oplus N_{SO(2)}$.
The (universal cover) of $SO(2) \times SO(3)$
is identified with the $U(1)_r \times SU(2)_R$
R-symmetry of the 4d field theory.
In summary,
the tangent bundle to 11d spacetime
restricted on $W_6$
decomposes as
\begin{equation} \label{TM11_decomp}
TM_{11} |_{W_6} = TM_4 \oplus T\Sigma_{g,n} \oplus N_{SO(2)} \oplus
N_{SO(3)} \ .
\end{equation}
The total space of the $N_{SO(2)}$ fibration over
$\Sigma_{g,n}$ is   the cotangent bundle
$T^* \Sigma_{g,n}$, and is hyper-K\"ahler.
The twisting of
$N_{SO(2)}$ over  $\Sigma_{g,n}$  implements a partial
topological twist
of the 6d $(2,0)$ $A_{N-1}$ theory
living on the stack. If $\hat{n}$ denotes the Chern root
of $N_{SO(2)}$, then
\begin{equation} \label{Chern_root}
\hat{n} = - \hat t + 2 \, c_1^{r} \ , \qquad
\int_{\Sigma_{g,n}} \hat t = \chi(\Sigma_{g,n})  \ ,
\end{equation}
where $c_1^r$ is the first Chern class
of   $U(1)_r$,
$\hat t$ is the Chern root of $T\Sigma_{g,n}$,
and $\chi(\Sigma_{g,n}) = 2(1-g)-n$
is the Euler characteristic of the punctured Riemann surface.
In order to specify the 4d theory, we must supplement each
puncture with appropriate data,
encoding the boundary conditions for
the 6d theory.
The puncture data is determined by 
the branching pattern
of the M5-branes
which governs the flavor symmetry
of the 4d theory.

From the point of view of M-theory, 
the combined system of the M5-brane stack and the 11d bulk 
enjoys a non-anomalous diffeomorphism invariance.
The total system is free from 
local anomalies in 11d due to a  cancellation between the 
anomaly generated by the chiral massless degrees of freedom
localized on $W_6$, and   anomaly inflow
from the  bulk.

The anomaly inflow from the bulk amounts to a
classical anomalous variation of the M-theory
effective action under 11d diffeomorphisms,
due to the presence of the 
 M5-brane stack. The latter    acts as
 a magnetic source for the M-theory four-form $G_4$
 with delta-function support on $W_6$, 
$dG_4 = 2\pi \, N \, \delta_{W_6}$.
In order to analyze   anomaly inflow in the supergravity
approximation we must smooth out the delta-function
singularity \cite{Freed:1998tg, Harvey:1998bx}. This is achieved by cutting out a small
tubular neighborhood of the M5-brane stack.
As a result, we are now considering M-theory
on a manifold with a boundary $M_{10} = \partial M_{11}$, which
is diffeomorphic
to an $S^4$ bundle over $W_6$.
The information about the original delta-function
source is translated into a smoothed-out
$\widetilde G_4$ flux,
\begin{equation} \label{tildeG4}
\frac{\widetilde G_4}{2\pi} = \frac{dC_3}{2\pi}
- df \wedge E_3^{(0)}
- f \, E_4 \ , \qquad
\int_{S^4} E_4= N \ .
\end{equation}
The quantity $f$ is a bump function that depends only
on the 
radial distance away from the M5-brane stack,
smoothly interpolating between $-1$ at the boundary $M_{10}$ and $0$
away from it.
The four-form $E_4$ is  
globally-defined,
closed, invariant under the action of the structure group
of $NW_6$, and can be written locally as $E_4 = dE_3^{(0)}$.
The integral of $E_4$ over the $S^4$ surrounding the
 stack measures the total magnetic charge~$N$
 of the M5-branes.

The anomalous variation
of the M-theory effective action 
is expressed as an integral over $M_{10}$
and is conveniently formulated in the 
framework of
descent,  
\begin{equation}
\frac{\delta S}{2\pi} =   \int_{M_{10}} \mathcal I_{10}^{(1)}   \ , \quad
\mathcal I_{12} = d\mathcal I_{11}^{ (0)}    \ , \quad
 \delta \mathcal I_{11}^{(0)} = d\mathcal I_{10}^{(1)}   \ .
\end{equation}
The formal quantity $\mathcal I_{12}$ is a twelve-form
characteristic class constructed from $E_4$
and given by
\begin{equation}
\mathcal I_{12} = - \frac 16 \, (E_4)^3
- E_4 \, I_8 \ .
\end{equation}
On the right-hand-side we suppressed wedge products for brevity,
and we introduced the eight-form class $I_8$,
which is defined in terms of the Pontryagin classes  
of $TM_{11}$ as
\begin{equation} \label{I8_def}
I_8 = \frac{1}{192} \Big[ p_1(TM_{11})^2 - 4 \, p_2(TM_{11}) \Big] \ .
\end{equation}
The inflow contribution to the anomaly polynomial
of the 4d  CFT is extracted
by integrating $\mathcal I_{12}$ over the total space of the $S^4$ bundle over~$\Sigma_{g,n}$, denoted $M_6$,
\begin{equation} \label{I6infl_integral}
\mathcal I_6^{\rm inf} = \int_{M_6} \mathcal I_{12}  \ , \qquad 
S^4 \hookrightarrow M_6 \rightarrow \Sigma_{g,n}   \ .
\end{equation}
Anomaly cancellation requires $\mathcal I_6^{\rm inf}$ to
cancel against the CFT anomaly, up to decoupling modes,
\begin{equation}
\mathcal I_6^{\rm inf} + \mathcal I_6^{\rm CFT}  
+ \mathcal I_6^{\rm decoup}  = 0 \ .
\end{equation}

To compute the integral in (\ref{I6infl_integral}),
we excise small disks around each puncture on $\Sigma_{g,n}$,
together with the $S^4$ fibers on top of them.
We thus obtain a   space $\widetilde M_6$,
which is an $S^4$ fibration over a smooth Riemann surface 
with $n$ boundaries.
We   replace the excised portions of $M_6$
with suitable  local geometries $X_6^\alpha$, with $\alpha = 1, \dots, n$,
glued smoothly to $\widetilde M_6$. 
This decomposition of $M_6$ translates to
\begin{eqnarray} \label{integral_decomp}
\textstyle \mathcal I_6^{\rm inf} &=& \textstyle
\int_{\widetilde M_6} \mathcal I_{12}
+ \sum_{\alpha = 1}^n 
\int_{  X_6^\alpha} \mathcal I_{12}    \nonumber \\
&\equiv& \textstyle \mathcal I_6^{\rm inf}(\Sigma_{g,n})
+ \sum_{\alpha = 1}^n \mathcal I_6^{\rm inf}(P_\alpha)  \ , \label{eq:i6}
\end{eqnarray}
where $P_\alpha$ denotes the $\alpha^{\rm th}$ puncture
on $\Sigma_{g,n}$.
We refer to $\mathcal I_6^{\rm inf}(\Sigma_{g,n})$
as the bulk contribution to $ \mathcal I_6^{\rm inf}$.

Each geometry $X_6^\alpha$ is  locally $S^2 \times X^\alpha_4$,
where the $S^2$ encodes the angular directions of  $N_{SO(3)}$,
while  $X^\alpha_4$ comprises   
 the directions 
of the excised disk,
together with the fibers of $N_{SO(2)}$ on top of it.
More precisely, $X_4^\alpha$ is the local space that
models  $T^*\Sigma_{g,n}$ in the vicinity of the
puncture $P_\alpha$.
Thus, the possible choices of $X^\alpha_4$
in M-theory
encode the puncture data.
The space $X^\alpha_4$ admits a $U(1)$ isometry,
which is identified with the 
$U(1)$ action on $N_{SO(2)}$
in the bulk of $T^*\Sigma_{g,n}$.

\section{Bulk contribution to inflow}

To write down the class $E_4$ on $\widetilde M_6$ it is convenient to
recall that $S^4$ can be realized as an 
$S^1_\phi \times S^2_\Omega$ fibration
over an interval. The subscript $\phi$ is a reminder that
we use the coordinate $\phi$ (with period $2\pi$)
to parametrize $S^1_\phi$. The label $\Omega$ is inserted for convenience,
to distinguish $S^2_\Omega$ from other two-spheres
discussed below.
The interval is parametrized with a coordinate $\mu \in [0,1]$.
At $\mu = 0$ the radius of $S^2_\Omega$ goes to zero,
while at $\mu = 1$   $S^1_\phi$     shrinks to zero.
The non-triviality of the $N_{SO(2)}$ bundle
is captured by $D\phi = d\phi - \mathcal A$, where $\mathcal A$ is a connection
with  field strength
$d\mathcal A = 2 \pi \, \hat n$, see (\ref{Chern_root}).
Using this notation,
the general $E_4$ reads
\begin{equation} \label{bulk_E4}
E_4= N \, \bigg[ d\gamma \wedge \frac{D\phi}{2\pi}
- \gamma \, \hat n \bigg] \wedge e_2^\Omega \ .
\end{equation}
The function $\gamma$ depends on $\mu$ only,
satisfies
$\gamma(0) = 0$, $\gamma(1) = 1$,
and has no zeros within the interval $(0,1)$, but is otherwise 
arbitrary.
The two-form
$e_2^\Omega$ is the closed, $SO(3)$-invariant 
completion of the volume form on $S^2_\Omega$,
normalized  to
integrate to $1$.
The overall normalization in (\ref{bulk_E4}) is fixed
by (\ref{tildeG4}).

The class $I_8$  on $\widetilde M_6$ is
obtained via the decomposition of $p_1(TM_{11})$,
$p_2(TM_{11})$ under  (\ref{TM11_decomp}),
using standard formulae for 
Pontryagin classes of direct sums of vector bundles.
Notice that $p_1(T\Sigma_{g,n}) = \hat t^2$,
$p_1(N_{SO(2)}) = \hat n^2$, while 
$p_1(N_{SO(3)}) = - 4 \, c_2^R$,
where $c_2^R$ is the second Chern class
of    $SU(2)_R$.
The only terms in $I_8$ that can contribute to the integral
over $\widetilde M_6$ are those linear in $\hat t$,
\begin{equation}
I_8 =  \frac {1}{48} \, \hat t \, c_1^r \, \Big[
4 \, (c_1^r)^2
+ 4 \, c_2^R
- p_1(TM_4) 
  \Big]
+ \cdots 
\end{equation}

We are now in a position to compute the integral of
$\mathcal I_{12}$ over $\widetilde M_6$.
To this end, it is useful to recall the Bott-Cattaneo formula \cite{bott1999integral}
$\int_{S^2_\Omega} (e_2^\Omega)^3 =  - c_2^R$.
The result reads
\begin{eqnarray} \label{bulk_result}
\mathcal I_6^{\text{inf}}(\Sigma_{g,n}) &=&\frac 12 \, N \, \chi(\Sigma_{g,n}) \, 
\bigg[ \frac { (c_1^r)^3 }{3}  - \frac{ c_1^r \, p_1(TM_4) }{12}
\bigg]  \nonumber \\
&-&   \frac 16  \, (4 \, N^3 - N) \, \chi(\Sigma_{g,n}) \, c_1^r \, c_2^R \ .
\end{eqnarray}
The quantity $\mathcal I_6^{\text{inf}}(\Sigma_{g,n})$ 
coincides with the dimensional reduction along $\Sigma_{g,n}$
of the inflow eight-form anomaly polynomial
for a stack of M5-branes \cite{Bah:2018gwc}.

\section{Puncture geometry and flux}

\begin{figure}[h]
\includegraphics[width=7.7cm]{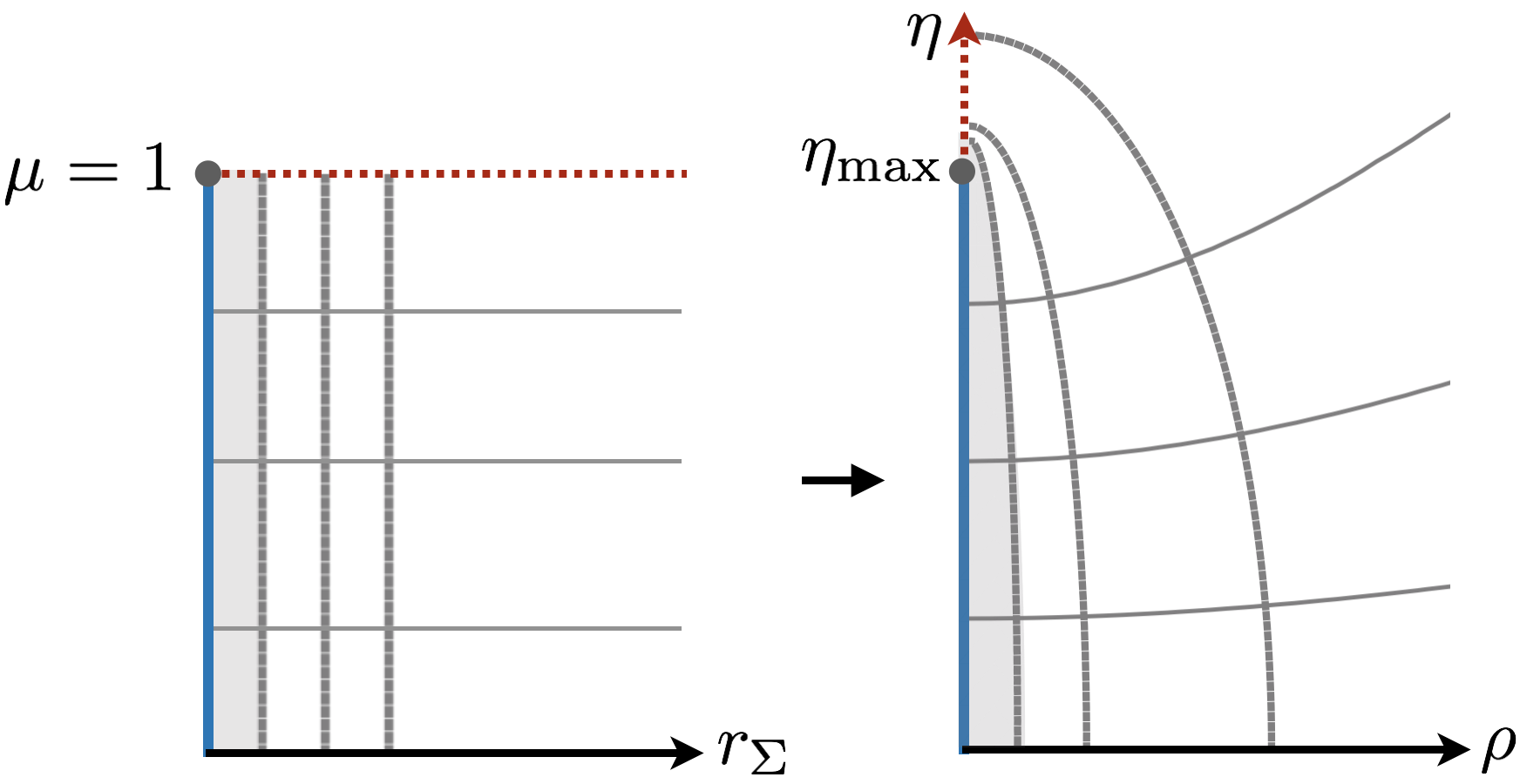}
\caption{The left plot shows the $(r_\Sigma, \mu)$ strip. The right plot shows the
$(\rho ,\eta)$ quadrant, including lines of constant $\mu$ and $r_\Sigma$. The near-puncture region is shaded with grey. \label{fig:1}}
\end{figure}

To understand $X_6^\alpha$, first consider a small disk around a generic point on $\Sigma_{g,n}$ with polar coordinates $(r_\Sigma,\beta)$.  The local geometry is an $S^1_\beta\times S^1_\phi\times S^2_\Omega$ fibration over the half-strip spanned by $r_\Sigma$ and the $\mu$ interval depicted in Figure \ref{fig:1}.    $S^2_\Omega$ shrinks along the boundary component at $\mu=0$ (the black line);   $S^1_\phi$ shrinks along $\mu=1$ (the dotted red line); and   $S^1_\beta$ shrinks along $r_\Sigma=0$ (the blue line).

We now map the $(r_\Sigma, \mu)$ half-strip
to a quadrant of $\mathbb R^2$ with coordinates
$(\rho, \eta)$. The qualitative features of this map 
are highlighted in Figure \ref{fig:1}.
The region $0 < \eta < \eta_{\rm max}$
on the $\eta$ axis corresponds to the $\mu$ interval at $r_\Sigma=0$,
while the region~$\eta > \eta_{\rm max}$~corresponds to
$r_\Sigma>0$~at~$\mu = 1$.

We define a new angle $\chi = \phi + \beta$,
and we regard the whole $X_4^\alpha$
as an  $S^1_\beta$ fibration over the 3d base
space spanned by $(\rho,\eta,\chi)$. 
We demand  that $S^1_\chi$ shrinks along the
$\eta$ axis \emph{in the base space},
so that we identify the base space with $\mathbb R^3$ with cylindrical coordinates
$(\rho,\eta,\chi)$.
The non-triviality of the $S^1_\beta$ fibration is 
captured~by
\begin{equation} \label{Dbeta}
D\beta = d\beta - L \, d\chi \ , \qquad
S^1_\beta \hookrightarrow X_4^\alpha \rightarrow \mathbb R^3 \ ,
\end{equation}
for $L$ a function of $\rho$, $\eta$.

The function $L$ is smooth in the interior of the
$(\rho, \eta)$ quadrant, but it approaches a discontinuous,
piecewise constant
function  of $\eta$ for $\rho \rightarrow 0$.
More precisely, we need 
$L = 1$ for $0 < \eta < \eta_{\rm max}$,
and $L = 0$ for $\eta > \eta_{\rm max}$.
This ensures that we 
reproduce the features of the previous description---that $S^1_\phi$ shrinks at $\eta > \eta_{\text{max}}$ and $S^1_\beta$ shrinks on $[0,\eta_{\text{max}}]$.
The discontinuity in   $L$ implies that the $S^1_\beta$ fibration has a monopole source
of charge $+1$ on the $\eta$ axis located at $\eta = \eta_{\text{max}}$.

So far we have done a
rewriting of the local geometry near a generic point on the Riemann surface,
\emph{i.e.}~a \emph{non-puncture}. 
Although in this case the local geometry is trivial,
the formulation in terms of the 
 $S^1_\beta$ fibration (\ref{Dbeta}) features a 
 monopole source where the $S^1_\beta$ fiber
 shrinks.

\begin{figure}[h!]
\includegraphics[width=3.3cm]{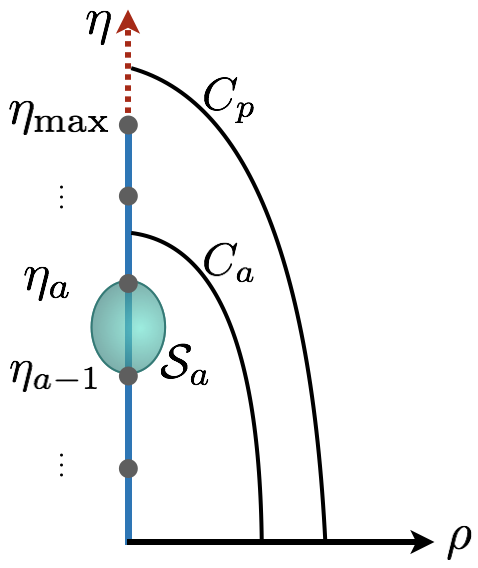}
\caption{  A generic profile of monopoles. The $C_{a}$ arcs form part of the four-cycle $\mathcal{C}_{a}$. The bubble   denotes the 
two-cycle $\mathcal S_a$, which is part of the four-cycle~$\mathcal{B}_a$. \label{fig:monopoles}}
\end{figure}

This setup lends itself to a natural generalization.
Consider 
a fibration as in (\ref{Dbeta}) with
$p$ monopoles labeled by $a=1,\dots,p$, located at $\eta=\eta_a$ and with
$\eta_p = \eta_{\rm max}$.
  This configuration is depicted in Figure \ref{fig:monopoles}. Denote the piecewise constant values of $L$ by
	\begin{equation}
	L = \ell_a\quad\text{for}\quad \eta_{a-1}<\eta<\eta_a \ ; \quad  \ell_{p+1}=0 \ . \label{eq:Ls}
	\end{equation}
The charge $k_a$ of each monopole is measured by 
\begin{equation} \label{eq:monopole_flux}
\int_{S^2_a} \frac{dD\beta}{2\pi} \equiv k_a = \ell_a - \ell_{a+1} \in \mathbb{Z}   \ ,
\end{equation}
for $S_a^2$ the 2-sphere surrounding the monopole in base space $\mathbb R^3$.
The $S^1_\beta$ circle shrinks at  each monopole.

Since the space $X_4^\alpha$ is a local model for
$T^*\Sigma_{g,n}$ in the neighborhood of the puncture $P_\alpha$,
its geometry is constrained.
In particular,
$k_a>0$ for all $a$, so that the $\ell_a$ are a sequence of 
decreasing integers. 
Furthermore, 
the local geometry near each monopole
is an ALF hyper-K\"ahler space,
modeled by a single-center Taub-NUT space
with charge $k_a$, denoted ${\rm TN}_{k_a}$.
This space has an 
$\mathbb R^4/\mathbb Z_{k_a}$
orbifold singularity which
can be resolved to yield a smooth
hyper-K\"ahler space $\widetilde{\rm TN}_{k_a}$.

Now we discuss $E_4$ in the geometry $X_6^\alpha$.
The most general form of $E_4$ compatible with 
the symmetries is
\begin{equation} \label{E4_puncture}
E_4 = d \Big(Y \, D\chi - W \, \widetilde {D\beta} \Big) \wedge e_2^\Omega \ ,
\quad
D\chi \equiv d\chi - \mathcal A \ ,
\end{equation}
where the gauging of $\chi$ with the connection $\mathcal A$
is inherited from $\phi$,
and $\widetilde {D\beta}$
denotes $D\beta$ as in (\ref{Dbeta})
with $d\chi \rightarrow D\chi$.
 The field strength $d\mathcal A$
in the puncture region only receives contributions from the
term $2 \, c_1^r$ in (\ref{Chern_root}).
The quantities $Y$, $W$ are functions of $\rho$, $\eta$
and are constrained by flux quantization of $E_4$.
Both $Y$ and $W$ must vanish 
on the $\rho$ axis at $\eta = 0$, because $S^2_\Omega$ shrinks there.

We start by defining the relevant cycles. 
For $a=1,\dots,p$ there is a four-cycle $\mathcal B_a$
consisting of the interval $[\eta_{a-1}, \eta_a]$
at $\rho = 0$,   $S^1_\beta$, and   $S^2_\Omega$.
For $a \ge 2$,  $S^1_\beta$ shrinks at the endpoints of 
$[\eta_{a-1}, \eta_a]$ and thus we also have a two-cycle
$\mathcal S_a$, depicted in Figure  \ref{fig:monopoles}.

Next, consider the arc $C_a$ connecting
a point on the $\rho$ axis to a point
within the $(\eta_{a},\eta_{a+1})$ interval, with $a = 1, \dots, p-1$,
as depicted in Figure \ref{fig:monopoles}. 
The arc $C_a$, together with  
 $S^2_\Omega$ and  the combination of $S^1_\chi$ and $S^1_\beta$
that shrinks along  $(\eta_{a}, \eta_{a+1})$,  
gives the four-cycle $\mathcal C_a$.
The arc $C_{p}$ in Figure \ref{fig:monopoles},
combined with  $S^1_\phi$  and $S^2_\Omega$,
gives a four-cycle $\mathcal C_{p}$ that is equivalent to the bulk $S^4$.

Supersymmetry requires the flux of $E_4$ through the $\mathcal{C}_a$ and $\mathcal{B}_a$ cycles to respectively carry the same sign.  We choose the orientations such that $\int_{\mathcal B_a} E_4$ and $\int_{\mathcal C_a} E_4$ are positive to be consistent with the conditions $\mathcal{C}_p \cong S^4$ and,
for the non-puncture,   $\mathcal{C}_1 \cong \mathcal{B}_1 \cong S^4$. 
One finds
\begin{equation}
 \int_{\mathcal B_a} E_4 = 
W(0, \eta_a) - W(0, \eta_{a-1}) \equiv w_a - w_{a-1} 
  \ , \label{eq:ba}
\end{equation}
such that $w_0 = 0$ and $\{w_a\}_{a=1}^p$ is an increasing sequence of positive 
integers.

The flux $\int_{\mathcal C_a} E_4$
equals 
$Y$ evaluated at the endpoint
of the $C_a$ arc on the $\eta$ axis.
Since the endpoint can be freely moved within $(\eta_a, \eta_{a+1})$,
$Y$ is piecewise constant along the $\eta$ axis,
and takes non-negative integer values,
\begin{equation}
Y(0,\eta) = y_a \in \mathbb Z_{\ge 0}   \quad
    \text{for}   \quad \eta_{a}<\eta<\eta_{a+1} \ .
\end{equation}
Although $Y$ is discontinuous along the $\eta$ axis,
$E_4$ must be continuous. This condition gives 
$y_{a} - y_{a-1} = w_a \, k_a$,
\begin{equation}
    y_a = \sum_{b=1}^a w_b \, k_b  \ , \qquad
    N = \sum_{a=1}^p w_a \, k_a \ , \label{eq:partition}
\end{equation}
where  $y_0 = 0$ and we used $\mathcal C_{p} \cong S^4$.
Continuity of $E_4$ thus implies  the partition of $N$
labeling a regular puncture.

For each non-trivial two-cycle in $X_6^\alpha$, we can turn on an additional contribution to $E_4$
of the form  $\omega \wedge F$,
for $\omega$ the
Poincar\'e dual of the two-cycle and $F$ the 
field strength of a background $U(1)$ connection
on $M_4$. One such two-cycle is 
$\mathcal S_a$
depicted in  Figure \ref{fig:monopoles}, 
with Poincar{\'e} dual denoted     $\omega_a$.
Additional two-cycles are introduced upon
resolving the orbifold singularities at the monopoles.
The resolved space $\widetilde {\rm TN}_{k_a}$
admits $k_a-1$ two-cycles, with Poincar\'e duals
$\{ \widehat{\omega}_{a,I} \}_{I=1}^{k_a-1}$.
Their intersection pairings  
 give the Cartan matrix 
$C^{\mathfrak{su}(k_a)}$
of
$\mathfrak{su}(k_a)$,
\begin{equation}
\int_{\widetilde {\rm TN}_{k_a}} \widehat \omega_{a,I} \wedge 
\widehat \omega_{a,J} = - C^{\mathfrak{su}(k_a)}_{IJ} \ .
\end{equation}
Including these additional terms, $E_4$ reads
\begin{eqnarray} \label{fullE4}
E_4 &=& d \Big(Y \, D\chi - W \, \widetilde {D\beta} \Big) \wedge e_2^\Omega
\nonumber \\
&+&    \sum_{a=2}^{p} \omega_a \wedge    \frac{F_a}{2\pi}
+   \sum_{a=1}^p \sum_{I=1}^{k_a-1} \widehat \omega_{a,I} \wedge   \frac{\widehat F_{a,I}}{2\pi} \ ,
\end{eqnarray}
where $F_a$ and $\widehat F_{a,I}$ are 4d field strengths.
(\ref{fullE4})  only captures the Cartan subgroup of 
 the full
4d flavor~group~$G_F$,
\begin{equation}
G_F = S\left[ \prod_{a=1}^p U(k_a) \right] \ . \label{eq:flavorsymm}
\end{equation}
  
Let us now discuss $I_8$ in the puncture geometry.
It is computed using the local decomposition
\begin{equation}
TM_{11} = TM_4 \oplus N_{SO(3)} \oplus TX_4^\alpha \ .
\end{equation}
The Pontryagin classes of $TX_4^\alpha$ are given in terms
of the Chern roots $\lambda_1$, $\lambda_2$ as
$p_1(TX_4^\alpha) = \lambda_1^2 + \lambda_2^2$,
$p_2(TX_4^\alpha) = \lambda_1^2  \lambda_2^2$.
To account for the
gauging of the angle $\chi$ in (\ref{E4_puncture}), the Chern roots are 
 shifted by $c_1^r$,
\begin{equation} \label{Chern_shift}
\lambda_1 \rightarrow \lambda_1 +  c_1^r \ , \qquad
\lambda_2  \rightarrow  \lambda_2 + c_1^r \ .
\end{equation}
The relevant terms of $I_8$ are
\begin{equation}
I_8 = \frac{1}{96} \Big[  4 \, (c_1^r)^2 + 4 \, c_2^R     - p_1(TM_4) \Big]
\, p_1(TX_4^\alpha)  + \cdots
\end{equation}
where $p_1(TX_4^\alpha)$ is taken
as the class before the shift (\ref{Chern_shift}).
The total $p_1(TX_4^\alpha)$ decomposes into a sum of
$p_1( \widetilde {\rm TN}_{k_a})$ terms,
which satisfy
$
\int_{\widetilde {\rm TN}_{k_a}}   p_1( \widetilde {\rm TN}_{k_a}) = 2 \, k_a 
$
 \cite{Gibbons:1979gd}.

\section{Inflow answer and cft comparison}

We now have the necessary components to compute $\mathcal I_6^{\rm inf}(P_\alpha)=
\int_{X_6^\alpha} \mathcal I_{12}$ in (\ref{eq:i6}). 
We use the standard parametrization of $\mathcal I_6$ for 4d $\mathcal N = 2$ SCFTs
\begin{eqnarray}
\mathcal I_6 &=&   (n_v-n_h)\left[ \frac{ (c_1^r)^3}{3} - \frac{c_1^r \, p_1(TM_4)}{12} \right] \nonumber \\
 &-& n_v \,  c_1^r \, c_2^R +\textstyle \sum_{G} k_{G}  \, c_1^r  \, c_2( G)\ , \label{eq:i6param}
\end{eqnarray} 
where $n_v$ and $n_h$ are the effective numbers of vector multiplets and hypermultiplets respectively;
$k_{G}$ is the flavor central charge of a factor $G$
of the 4d flavor   group.

\newpage

A direct computation of the integrals yields
\begin{align}
& (n_v-n_h)^{\text{inf}}(P_\alpha)  =   \textstyle\frac{1}{2} \sum_{a=1}^p N_a k_a\ , \label{eq:answer1}\\
& n_v^{\text{inf}}(P_\alpha)   =  \textstyle{ \sum_{a=1}^p} \Big[  \frac{2}{3} \, \ell_a^2 \,(w_a^3 - w_{a-1}^3) - \frac{1}{6} N_a k_a  \nonumber \\
& \qquad \qquad \quad \quad \ \
+ \ell_a \, (N_a-  w_a \ell_a) \, (w_a^2 - w_{a-1}^2) \Big] \ , \label{eq:answer2}\\
& k_{SU(k_a)}^{\text{inf}}   =   -2N_a  \ ,
\quad N_a \equiv \textstyle \sum_{b=1}^a (w_b - w_{b-1}) \,  \ell_b  \ .
 \label{eq:answer3}
\end{align}
Note that there is an enhancement of the $k_a-1$ Cartan components to the second Chern class of the full non-Abelian $SU(k_a)$ factor in (\ref{eq:flavorsymm}).

The partition of $N$ in (\ref{eq:partition}) defines a Young diagram
with rows $\{ \widetilde \ell_i \}_{i=1}^{w_p}$,
where $\widetilde \ell_i = \ell_a$ for $w_{a-1}+1 \le i \le w_a$.
We define 
$\widetilde{k}_i = \widetilde{\ell}_i - \widetilde{\ell}_{i+1}$
and $\widetilde{N}_i = \sum_{j=1}^i \widetilde{\ell}_j$.
It follows  that (\ref{eq:answer1})-(\ref{eq:answer2}) are equivalently written as
\begin{eqnarray}
(n_v-n_h)^{\text{inf}}(P_\alpha) &=& \textstyle \frac{1}{2} \sum_{i=1}^{w_p} \widetilde{N}_i  \, \widetilde{k}_i\ , \label{eq:answer10}\\
n_v^{\text{inf}}(P_\alpha) &=& \textstyle \sum_{i=1}^{w_p } (N^2-\widetilde{N}_i^2) + \frac{1}{2} N^2   \label{eq:answer20}  \ .
\end{eqnarray}
We can also read off
$n_{v,h}^{\rm inf}(\Sigma_{g,n})$
from (\ref{bulk_result}),
\begin{eqnarray}
(n_v - n_h)^{\rm inf}(\Sigma_{g,n}) &=& 
\tfrac 12 \, N \, \chi(\Sigma_{g,n})
 \ , \\
n_h^{\rm inf}(\Sigma_{g,n}) &=& \tfrac 16 \, (4 \, N^3 - N)  \, \chi(\Sigma_{g,n}) \ . 
\end{eqnarray}
According to (\ref{integral_decomp}), the total $n_v^{\rm inf}$, $n_h^{\rm inf}$ are   
\begin{equation}
n_{v,h}^{\rm inf} = n_{v,h}^{\rm inf} (\Sigma_{g,n}) 
+ \textstyle  \sum_{\alpha = 1}^n
n_{v,h}^{\rm inf} (P_\alpha) \ .
\end{equation}
These quantities can now be compared to the
known CFT answers \cite{Chacaltana:2012zy},
as presented in \cite{Bah:2018gwc}. 
We find
\begin{gather}
n_v^{\rm inf} + n_v^{\rm CFT}  = \tfrac 12 \, \chi(\Sigma_{g,0}) \ ,  \qquad
n_h^{\rm inf} + n_h^{\rm CFT} =0 \ , \\
k_{ SU(k_a)}^{\text{inf}}  + k_{ SU(k_a)}^{\text{CFT}}   =0 \ .
\end{gather}
The inflow and CFT contributions cancel,
up to 
minus the anomaly of a free 6d $(2,0)$ tensor multiplet reduced
on a genus-$g$   Riemann surface $\Sigma_{g,0}$ \emph{with no punctures}.
We identify this free tensor multiplet with the center-of-mass
mode of the M5-brane stack. Our results show that  this mode  is 
insensitive to the presence of punctures.

\section{Conclusion and applications to holography} 

In this letter we provided a first principles
derivation of the anomaly polynomials of 4d $\mathcal N = 2$
$A_{N-1}$ class $\mathcal S$ theories with arbitrary regular punctures,
using anomaly inflow in the corresponding M-theory setup
with $N$ M5-branes wrapping a punctured Riemann surface.

In our approach, the puncture data are entirely specified
by the topological properties
of the 11d geometry and $G_4$ flux
in the vicinity of the  puncture.
Remarkably, the anomaly inflow cancels exactly  
the known anomalies of the 4d SCFTs,
up to the contribution of the
center-of-mass free tensor multiplet on the M5-brane stack.

Our method for analyzing $\mathcal N =2$ regular punctures
is   generalizable to irregular punctures and setups with less supersymmetry.
Many interesting QFTs can be realized via branes 
probing   geometries
in string theory and M-theory.
In such cases,  inflow 
can be a robust tool to compute 
anomalies, and therefore provides a handle on non-perturbative
aspects of these QFTs.

We conclude with a discussion of applications to holography.
An important motivation for our analysis of the local
puncture geometry and $E_4$ flux  
comes from the    holographic M-theory duals of $\mathcal N = 2$ and $\mathcal N = 1$
class
$\mathcal S$ theories with punctures \cite{Gaiotto:2009gz, Bah:2015fwa}.
In particular, the fibration in (\ref{Dbeta})
is related to and inspired by   the B\"acklund transform
of \cite{Gaiotto:2009gz}.
The solutions are warped products of $AdS_5$ with an internal
space $M_6^{\rm hol}$
with four-form flux $G_4^{\rm hol}$.

We observe that the topological properties of 
$M_6^{\rm hol}$ in \cite{Gaiotto:2009gz}
are the same as those of $M_6$ in (\ref{I6infl_integral}).
Furthermore, 
\begin{equation} \label{holo_flux}
\frac{ G_4^\text{hol}  }{2\pi}  =   \overline E_4  \quad
\text{in cohomology} \ ,  
\end{equation}
where $\overline E_4$ is $E_4$ with all 4d connections
turned off  
and $G_4^\text{hol}$  is the four-form flux of  \cite{Gaiotto:2009gz}.
In the bulk of $\Sigma_{g,n}$ $\overline E_4  = S^4$,
but $\overline E_4$ is non-trivial in the puncture geometry
and encodes the puncture labelling.

Kaluza-Klein reduction of 11d supergravity on $M_6^{\rm hol}$
yields a 5d gauged supergravity model with an $AdS_5$ vacuum.
The full reduction ansatz requires a  $G_4^{\rm hol}$ that
captures the fluctuations of the
$AdS_5$ gauge fields beyond the linearized level.
$E_4$ is a natural
candidate
for constructing such an ansatz \cite{Harvey:1998bx}.

In the solutions of \cite{Gaiotto:2009gz} the \emph{classical} objects $M_6^{\rm hol}$,   $G_4^\text{hol}$ 
provide the 
 \emph{exact} topological data of $M_6$,  $\overline E_4$
to all orders in $N$.
This data determines the $E_4$ and $I_8$ needed to 
carry out the inflow procedure,  which
(subtracting the $\mathcal O(1)$ contribution of decoupling modes)
yields the exact anomaly coefficients
of the dual SCFT. This route to the exact $a$ and $c$
central charges bypasses
a computation with the $AdS_5$ effective action,
which would require a detailed
knowledge of higher-derivative corrections.

An interesting question is whether (\ref{holo_flux})
extends to  more general $AdS_5$ solutions 
in M-theory, with varying amount of supersymmetry.
If so, we may use inflow and classical data of the supergravity solution
 to access exact
anomaly coefficients,  
providing a 
systematic way 
to compute
quantum corrections in $AdS_5$.

\begin{acknowledgments}
\emph{Acknowledgments.} We are grateful to  J.~Distler,  T.~Dumitrescu, K.~Intriligator, J.~Kaplan, C.~Lawrie, G.~Moore, S.~Sch\"afer-Nameki,
J.~Song, Y.~Tachikawa,  A.~Tomasiello, and Y.~Wang
 for interesting
conversations and correspondence.
The work of EN is supported in part by DOE grant DE-SC0009919, and a UC President's Dissertation Fellowship.
The work of IB and FB is   supported in part by NSF grant PHY-1820784.
Part of this work was performed at the Aspen Center for Physics, which is supported by NSF grant PHY-1607611.

\end{acknowledgments}

\bibliography{refs}

\begin{thebibliography}{12}%
\makeatletter
\providecommand \@ifxundefined [1]{%
 \@ifx{#1\undefined}
}%
\providecommand \@ifnum [1]{%
 \ifnum #1\expandafter \@firstoftwo
 \else \expandafter \@secondoftwo
 \fi
}%
\providecommand \@ifx [1]{%
 \ifx #1\expandafter \@firstoftwo
 \else \expandafter \@secondoftwo
 \fi
}%
\providecommand \natexlab [1]{#1}%
\providecommand \enquote  [1]{``#1''}%
\providecommand \bibnamefont  [1]{#1}%
\providecommand \bibfnamefont [1]{#1}%
\providecommand \citenamefont [1]{#1}%
\providecommand \href@noop [0]{\@secondoftwo}%
\providecommand \href [0]{\begingroup \@sanitize@url \@href}%
\providecommand \@href[1]{\@@startlink{#1}\@@href}%
\providecommand \@@href[1]{\endgroup#1\@@endlink}%
\providecommand \@sanitize@url [0]{\catcode `\\12\catcode `\$12\catcode
  `\&12\catcode `\#12\catcode `\^12\catcode `\_12\catcode `\%12\relax}%
\providecommand \@@startlink[1]{}%
\providecommand \@@endlink[0]{}%
\providecommand \url  [0]{\begingroup\@sanitize@url \@url }%
\providecommand \@url [1]{\endgroup\@href {#1}{\urlprefix }}%
\providecommand \urlprefix  [0]{URL }%
\providecommand \Eprint [0]{\href }%
\providecommand \doibase [0]{http://dx.doi.org/}%
\providecommand \selectlanguage [0]{\@gobble}%
\providecommand \bibinfo  [0]{\@secondoftwo}%
\providecommand \bibfield  [0]{\@secondoftwo}%
\providecommand \translation [1]{[#1]}%
\providecommand \BibitemOpen [0]{}%
\providecommand \bibitemStop [0]{}%
\providecommand \bibitemNoStop [0]{.\EOS\space}%
\providecommand \EOS [0]{\spacefactor3000\relax}%
\providecommand \BibitemShut  [1]{\csname bibitem#1\endcsname}%
\let\auto@bib@innerbib\@empty
\bibitem [{\citenamefont {Gaiotto}(2009)}]{Gaiotto:2009we}%
  \BibitemOpen
  \bibfield  {author} {\bibinfo {author} {\bibfnamefont {D.}~\bibnamefont
  {Gaiotto}},\ }\href@noop {} {\  (\bibinfo {year} {2009})},\ \Eprint
  {http://arxiv.org/abs/arXiv:0904.2715} {arXiv:arXiv:0904.2715 [hep-th]}
  \BibitemShut {NoStop}%
\bibitem [{\citenamefont {Gaiotto}\ \emph {et~al.}(2009)\citenamefont
  {Gaiotto}, \citenamefont {Moore},\ and\ \citenamefont
  {Neitzke}}]{Gaiotto:2009hg}%
  \BibitemOpen
  \bibfield  {author} {\bibinfo {author} {\bibfnamefont {D.}~\bibnamefont
  {Gaiotto}}, \bibinfo {author} {\bibfnamefont {G.~W.}\ \bibnamefont {Moore}},
  \ and\ \bibinfo {author} {\bibfnamefont {A.}~\bibnamefont {Neitzke}},\
  }\href@noop {} {\  (\bibinfo {year} {2009})},\ \Eprint
  {http://arxiv.org/abs/0907.3987} {arXiv:0907.3987 [hep-th]} \BibitemShut
  {NoStop}%
\bibitem [{\citenamefont {Gaiotto}\ and\ \citenamefont
  {Maldacena}(2012)}]{Gaiotto:2009gz}%
  \BibitemOpen
  \bibfield  {author} {\bibinfo {author} {\bibfnamefont {D.}~\bibnamefont
  {Gaiotto}}\ and\ \bibinfo {author} {\bibfnamefont {J.}~\bibnamefont
  {Maldacena}},\ }\href {\doibase 10.1007/JHEP10(2012)189} {\bibfield
  {journal} {\bibinfo  {journal} {JHEP}\ }\textbf {\bibinfo {volume} {10}},\
  \bibinfo {pages} {189} (\bibinfo {year} {2012})},\ \Eprint
  {http://arxiv.org/abs/0904.4466} {arXiv:0904.4466 [hep-th]} \BibitemShut
  {NoStop}%
\bibitem [{\citenamefont {Chacaltana}\ \emph {et~al.}(2013)\citenamefont
  {Chacaltana}, \citenamefont {Distler},\ and\ \citenamefont
  {Tachikawa}}]{Chacaltana:2012zy}%
  \BibitemOpen
  \bibfield  {author} {\bibinfo {author} {\bibfnamefont {O.}~\bibnamefont
  {Chacaltana}}, \bibinfo {author} {\bibfnamefont {J.}~\bibnamefont {Distler}},
  \ and\ \bibinfo {author} {\bibfnamefont {Y.}~\bibnamefont {Tachikawa}},\
  }\href {\doibase 10.1142/S0217751X1340006X} {\bibfield  {journal} {\bibinfo
  {journal} {Int. J. Mod. Phys.}\ }\textbf {\bibinfo {volume} {A28}},\ \bibinfo
  {pages} {1340006} (\bibinfo {year} {2013})},\ \Eprint
  {http://arxiv.org/abs/1203.2930} {arXiv:1203.2930 [hep-th]} \BibitemShut
  {NoStop}%
\bibitem [{\citenamefont {Tachikawa}(2015)}]{Tachikawa:2015bga}%
  \BibitemOpen
  \bibfield  {author} {\bibinfo {author} {\bibfnamefont {Y.}~\bibnamefont
  {Tachikawa}},\ }\href {\doibase 10.1093/ptep/ptv098} {\bibfield  {journal}
  {\bibinfo  {journal} {PTEP}\ }\textbf {\bibinfo {volume} {2015}},\ \bibinfo
  {pages} {11B102} (\bibinfo {year} {2015})},\ \Eprint
  {http://arxiv.org/abs/1504.01481} {arXiv:1504.01481 [hep-th]} \BibitemShut
  {NoStop}%
\bibitem [{\citenamefont {Bah}\ and\ \citenamefont
  {Nardoni}(2018)}]{Bah:2018gwc}%
  \BibitemOpen
  \bibfield  {author} {\bibinfo {author} {\bibfnamefont {I.}~\bibnamefont
  {Bah}}\ and\ \bibinfo {author} {\bibfnamefont {E.}~\bibnamefont {Nardoni}},\
  }\href@noop {} {\  (\bibinfo {year} {2018})},\ \Eprint
  {http://arxiv.org/abs/1803.00136} {arXiv:1803.00136 [hep-th]} \BibitemShut
  {NoStop}%
\bibitem [{\citenamefont {Bah}\ \emph {et~al.}()\citenamefont {Bah},
  \citenamefont {Bonetti}, \citenamefont {Minasian},\ and\ \citenamefont
  {Nardoni}}]{Bah:2018bn}%
  \BibitemOpen
  \bibfield  {author} {\bibinfo {author} {\bibfnamefont {I.}~\bibnamefont
  {Bah}}, \bibinfo {author} {\bibfnamefont {F.}~\bibnamefont {Bonetti}},
  \bibinfo {author} {\bibfnamefont {R.}~\bibnamefont {Minasian}}, \ and\
  \bibinfo {author} {\bibfnamefont {E.}~\bibnamefont {Nardoni}},\ }\href@noop
  {} {\bibinfo  {journal} {To appear}\ }\BibitemShut {NoStop}%
\bibitem [{\citenamefont {Freed}\ \emph {et~al.}(1998)\citenamefont {Freed},
  \citenamefont {Harvey}, \citenamefont {Minasian},\ and\ \citenamefont
  {Moore}}]{Freed:1998tg}%
  \BibitemOpen
\bibfield  {journal} {  }\bibfield  {author} {\bibinfo {author} {\bibfnamefont
  {D.}~\bibnamefont {Freed}}, \bibinfo {author} {\bibfnamefont {J.~A.}\
  \bibnamefont {Harvey}}, \bibinfo {author} {\bibfnamefont {R.}~\bibnamefont
  {Minasian}}, \ and\ \bibinfo {author} {\bibfnamefont {G.~W.}\ \bibnamefont
  {Moore}},\ }\href {\doibase 10.4310/ATMP.1998.v2.n3.a8} {\bibfield  {journal}
  {\bibinfo  {journal} {Adv. Theor. Math. Phys.}\ }\textbf {\bibinfo {volume}
  {2}},\ \bibinfo {pages} {601} (\bibinfo {year} {1998})},\ \Eprint
  {http://arxiv.org/abs/hep-th/9803205} {arXiv:hep-th/9803205 [hep-th]}
  \BibitemShut {NoStop}%
\bibitem [{\citenamefont {Harvey}\ \emph {et~al.}(1998)\citenamefont {Harvey},
  \citenamefont {Minasian},\ and\ \citenamefont {Moore}}]{Harvey:1998bx}%
  \BibitemOpen
  \bibfield  {author} {\bibinfo {author} {\bibfnamefont {J.~A.}\ \bibnamefont
  {Harvey}}, \bibinfo {author} {\bibfnamefont {R.}~\bibnamefont {Minasian}}, \
  and\ \bibinfo {author} {\bibfnamefont {G.~W.}\ \bibnamefont {Moore}},\ }\href
  {\doibase 10.1088/1126-6708/1998/09/004} {\bibfield  {journal} {\bibinfo
  {journal} {JHEP}\ }\textbf {\bibinfo {volume} {09}},\ \bibinfo {pages} {004}
  (\bibinfo {year} {1998})},\ \Eprint {http://arxiv.org/abs/hep-th/9808060}
  {arXiv:hep-th/9808060 [hep-th]} \BibitemShut {NoStop}%
\bibitem [{\citenamefont {Bott}\ and\ \citenamefont
  {Cattaneo}(1998)}]{bott1999integral}%
  \BibitemOpen
  \bibfield  {author} {\bibinfo {author} {\bibfnamefont {R.}~\bibnamefont
  {Bott}}\ and\ \bibinfo {author} {\bibfnamefont {A.~S.}\ \bibnamefont
  {Cattaneo}},\ }\href@noop {} {\bibfield  {journal} {\bibinfo  {journal} {J.
  Differential Geom}\ }\textbf {\bibinfo {volume} {48}},\ \bibinfo {pages} {91}
  (\bibinfo {year} {1998})},\ \Eprint {http://arxiv.org/abs/dg-ga/9710001}
  {arXiv:dg-ga/9710001 [dg-ga]} \BibitemShut {NoStop}%
\bibitem [{\citenamefont {Gibbons}\ \emph {et~al.}(1979)\citenamefont
  {Gibbons}, \citenamefont {Pope},\ and\ \citenamefont
  {Romer}}]{Gibbons:1979gd}%
  \BibitemOpen
  \bibfield  {author} {\bibinfo {author} {\bibfnamefont {G.~W.}\ \bibnamefont
  {Gibbons}}, \bibinfo {author} {\bibfnamefont {C.~N.}\ \bibnamefont {Pope}}, \
  and\ \bibinfo {author} {\bibfnamefont {H.}~\bibnamefont {Romer}},\ }\href
  {\doibase 10.1016/0550-3213(79)90109-3} {\bibfield  {journal} {\bibinfo
  {journal} {Nucl. Phys.}\ }\textbf {\bibinfo {volume} {B157}},\ \bibinfo
  {pages} {377} (\bibinfo {year} {1979})}\BibitemShut {NoStop}%
\bibitem [{\citenamefont {Bah}(2015)}]{Bah:2015fwa}%
  \BibitemOpen
  \bibfield  {author} {\bibinfo {author} {\bibfnamefont {I.}~\bibnamefont
  {Bah}},\ }\href {\doibase 10.1007/JHEP09(2015)163} {\bibfield  {journal}
  {\bibinfo  {journal} {JHEP}\ }\textbf {\bibinfo {volume} {09}},\ \bibinfo
  {pages} {163} (\bibinfo {year} {2015})},\ \Eprint
  {http://arxiv.org/abs/1501.06072} {arXiv:1501.06072 [hep-th]} \BibitemShut
  {NoStop}%
\end{thebibliography}%

\end{document}